\documentclass[preprint,showpacs, superscriptaddress, preprintnumbers,amsmath,amssymb]{revtex4}

\usepackage{booktabs}
\usepackage{color,graphicx}
\usepackage{amsmath}
\usepackage{dcolumn}
\usepackage{CJK}                 
\usepackage{bm}                  
\usepackage{soul}
\usepackage{enumerate}

\usepackage[dvipdfm,  
            pdfstartview=FitH,
            CJKbookmarks=true,
            bookmarksnumbered=true,
            bookmarksopen=true,
            colorlinks, 
            pdfborder=001,   
            linkcolor=blue,
            anchorcolor=blue,
            citecolor=blue
            ]{hyperref}

\usepackage{float}

\begin{document}

\title{Two-dimensional collective Hamiltonian for chiral and wobbling modes II: Electromagnetic transitions}

\author{X. H. Wu}
\affiliation{State Key Laboratory of Nuclear Physics and Technology,
School of Physics, Peking University, Beijing 100871, China}

\author{Q. B. Chen}
\affiliation{Physik-Department, Technische Universit\"{a}t M\"{u}nchen,
D-85747 Garching, Germany}

\author{P. W. Zhao}
\affiliation{State Key Laboratory of Nuclear Physics and Technology,
School of Physics, Peking University, Beijing 100871, China}

\author{S. Q. Zhang}
\email{sqzhang@pku.edu.cn}
\affiliation{State Key Laboratory of Nuclear Physics and Technology,
School of Physics, Peking University, Beijing 100871, China}

\author{J. Meng}
\email{mengj@pku.edu.cn}
\affiliation{State Key Laboratory of Nuclear Physics and Technology,
School of Physics, Peking University, Beijing 100871, China}

\affiliation{Yukawa Institute for Theoretical Physics,
Kyoto University, Kyoto 606-8502, Japan}

\begin{abstract}
  The intraband electromagnetic transitions in the framework of collective Hamiltonian for chiral and wobbling modes are calculated. By going beyond the mean field approximation on the orientations of rotational axis, the collective Hamiltonian provides the descriptions on both yrast band and collective excitation bands. For a system with one $h_{11/2}$ proton particle and one $h_{11/2}$ neutron hole coupled to a triaxial rotor ($\gamma=-30^\circ$), the intraband electromagnetic transitions given by the one-dimensional and two-dimensional collective Hamiltonian are compared to the results by the tilted axis cranking approach and particle rotor model. Compared with the tilted axis cranking approach, the electromagnetic transitions given by the collective Hamiltonian have a better agreement with those by the particle rotor model, due to the consideration of the quantum fluctuations.
\end{abstract}

\date{\today}


\maketitle


\section{Introduction}

In the first paper of this series \cite{Chen2016Phys.Rev.C44301}, the two-dimensional collective Hamiltonian method based on the titled axis cranking (TAC) approach has been developed to describe the nuclear chirality \cite{Frauendorf1997Nucl.Phys.A131} and wobbling motion \cite{Bohr1975}, both of which provide direct evidences for the existence of nuclear triaxiality.
The chirality in nuclear physics was first predicted by Frauendorf and Meng in 1997 \cite{Frauendorf1997Nucl.Phys.A131}, which stimulates lots of experimental efforts and more than 60 candidate chiral bands reported in the $A\sim80$, 100, 130, and 190 mass regions. For recent reviews and detailed data tables, see Refs.~\cite{Meng2010J.Phys.G64025, Meng2014Int.J.Mod.Phys.E1430016, Bark2014Int.J.Mod.Phys.E1461001, Meng2016Phys.Scr.53008, Raduta2016Prog.Part.Nucl.Phys.241, Frauendorf2018Phys.Scripta43003} and \cite{Xiong2018AtomicDataandNuclearDataTables}. The wobbling motion was originally suggested by Bohr and Mottelson in the 1970s \cite{Bohr1975}, and has been observed in the $A\sim 160$ \cite{Odegard2001Phys.Rev.Lett.5866, Jensen2002Phys.Rev.Lett.142503, Amro2003Phys.Lett.B197, Schoenwasser2003Phys.Lett.B9, Bringel2005Eur.Phys.J.A167, Hartley2009Phys.Rev.C41304}, 130 \cite{Matta2015Phys.Rev.Lett.82501, BiswasarXiv:1608.07840}, and 100 \cite{Zhu2009Int.J.Mod.Phys.E1717, Luo2013Proceedingsoftheinternationalsymposium2013} mass regions.

Theoretically, the nuclear chirality and wobbling motion have been extensively investigated with the particle rotor model (PRM) \cite{Frauendorf1997Nucl.Phys.A131, Bohr1975, Peng2003Phys.Rev.C44324, Koike2004Phys.Rev.Lett.172502, Tanabe2006Phys.Rev.C34305, Zhang2007Phys.Rev.C44307, Higashiyama2007Eur.Phys.J.A355, Tanabe2008Phys.Rev.C64318, Qi2009Phys.Lett.B175, Chen2010Phys.Rev.C67302, Lawrie2010Phys.Lett.B66, Rohozinski2011Eur.Phys.J.A90, Shirinda2012Eur.Phys.J.A118, Zhang2016Chin.Phys.C24102, Hamamoto2002Phys.Rev.C44305, Hamamoto2003Phys.Rev.C34312, Frauendorf2014Phys.Rev.C14322, Shi2015Chin.Phys.C54105, Tanabe2017Phys.Rev.C64315, Budaca2018Phys.Rev.C24302, Chen2018Phys.Rev.C41303, Chen2018Phys.Lett.B744, Streck2018arXiv:1805.03633, Budaca2018Phys.Rev.C14303} and the tilted axis cranking (TAC) approaches based on either the Woods-Saxon mean field \cite{Matta2015Phys.Rev.Lett.82501, Dimitrov2000Phys.Rev.Lett.5732} or more fundamental density functional theories \cite{Olbratowski2004Phys.Rev.Lett.52501, Olbratowski2006Phys.Rev.C54308, Kuti2014Phys.Rev.Lett.32501, Zhao2017Phys.Lett.B1, Petrache2018Phys.Rev.C41304}. Other approaches include the boson expansion approaches \cite{Ganev2009Phys.Rev.C44322, Brant2004Phys.Rev.C17304, Raduta2014J.Phys.G:Nucl.Part.Phys.35105, Raduta2016J.Phys.G:Nucl.Part.Phys.95107}, the pair truncated shell model \cite{Higashiyama2005Phys.Rev.C24315} and the projected shell model \cite{Bhat2014Phys.Lett.B218, Chen2017Phys.Rev.C51303, Shimada2018Phys.Rev.C24318, Shimada2018Phys.Rev.C24319, Chen2018Phys.Lett.B211}.
The TAC approach, based on mean-field approximation, provides a clear picture for the chirality and wobbling motion in terms of the orientation of the angular momentum vector relative to the density distribution. To describe the chiral and wobbling excitations beyond the mean-field, the random phase approximation was developed on top of the TAC solutions \cite{Mukhopadhyay2007Phys.Rev.Lett.172501, Almehed2011Phys.Rev.C54308, Shimizu1995Nucl.Phys.A559, Matsuzaki2002Phys.Rev.C41303, Matsuzaki2004Eur.Phys.J.A189, Matsuzaki2004Phys.Rev.C34325, Matsuzaki2004Phys.Rev.C64317, Shimizu2005Phys.Rev.C14306, Shimizu2008Phys.Rev.C24319, Shoji2009Prog.Theor.Phys.319, Frauendorf2015Phys.Rev.C64306}. Alternatively, the collective Hamiltonian based on the TAC solutions is proved to be very successful \cite{Chen2016Phys.Rev.C44301, Chen2013Phys.Rev.C24314, Chen2014Phys.Rev.C44306, Chen2016Phys.Rev.C54308}. Particularly, the collective Hamiltonian method is promising to unify the description of both quantum tunneling and vibrations.

In previous works \cite{Chen2016Phys.Rev.C44301, Chen2013Phys.Rev.C24314}, the one- and two-dimensional collective Hamiltonian (1DCH and 2DCH) were constructed and applied to investigate the chirality of the system with one $h_{11/2}$ proton particle and one $h_{11/2}$ neutron hole coupled to a triaxial ($\gamma=-30^\circ$) rotor. It is found that the chiral symmetry is restored in the collective Hamiltonian solutions, which are in agreement with the energy spectra for chiral doublet bands given by the PRM \cite{Frauendorf1997Nucl.Phys.A131}. Similar successes have been achieved in describing the wobbling motions in the simple, longitudinal, and transverse wobblers \cite{Chen2014Phys.Rev.C44306} and in the nucleus $^{135}$Pr \cite{Chen2016Phys.Rev.C54308}. Moreover, more excitation modes appear in the framework of the 2DCH, since both the broken chiral and signature symmetries are restored \cite{Chen2016Phys.Rev.C44301}.

Besides the energy spectra, the electromagnetic (EM) transition properties are important observables in identifying the nuclear chirality or wobbling motion. Based on the model with the configuration $\pi(1h_{11/2}) \otimes \nu(1h_{11/2})^{-1}$ and $\gamma=-30^\circ$, the criteria for ideal nuclear chirality are \cite{Wang2007Chin.Phys.Lett.664, Meng2010J.Phys.G64025, Meng2016Phys.Scr.53008}: (\romannumeral1) the near degeneracy of doublet bands; (\romannumeral2) the spin independence of $S(I)$; (\romannumeral3) the similar spin alignments; (\romannumeral4) the $B(M1)$ values as well as the resulting $B(M1)/B(E2)$ ratios present the odd-even staggering behavior; (\romannumeral5) the doublet bands have similar intraband $M1$ and $E2$ transition strengthes;  (\romannumeral6) the interband $E2$ transitions vanish at high spin region.
For wobbling motion, one of the most important features is that the interband EM transitions with $\Delta I=1$ between the wobbling bands are dominated by $E2$ rather than by $M1$ \cite{Bohr1975, Odegard2001Phys.Rev.Lett.5866, Hamamoto2002Phys.Rev.C44305, Frauendorf2014Phys.Rev.C14322, Frauendorf2015Phys.Rev.C64306}.

In this work, the collective Hamiltonian in previous works \cite{Chen2016Phys.Rev.C44301, Chen2013Phys.Rev.C24314, Chen2014Phys.Rev.C44306, Chen2016Phys.Rev.C54308} is extended to calculate the intraband EM transition probabilities and compared with those given by the TAC and PRM.
The paper is organized as follows. In Sec. \ref{theory}, the frameworks of the 1DCH and 2DCH are briefly introduced, and the formulae for the intraband EM transition probabilities are given. The numerical details are given in Sec. \ref{numerical}. In Sec. \ref{results}, the calculated results are presented and compared with the TAC and PRM. Finally, a summary is given in Sec. \ref{summary}.


\section{Theoretical framework} \label{theory}

\subsection{Collective Hamiltonian}

The collective Hamiltonian can be derived, for examples, by the generator coordinate method (GCM) \cite{Ring1980}, the adiabatic time-dependent Hartree-Fock (ATDHF) method \cite{Ring1980, Baranger1968Nucl.Phys.A241}, or the adiabatic self-consistent collective coordinate (ASCC) method \cite{Matsuo2000Prog.Theor.Phys.959, Matsuyanagi2010J.Phys.G64018}.

The orientations of the rotational axis in a triaxial nucleus can be parametrized by the polar and the azimuth angles $(\theta,\varphi)$. These two angles are chosen as the collective coordinates to describe the chiral and wobbling modes in the collective Hamiltonian method. Based on the TAC approach, the collective Hamiltonian of the azimuth angle $\varphi$ (1DCH) \cite{Chen2013Phys.Rev.C24314, Chen2014Phys.Rev.C44306, Chen2016Phys.Rev.C54308} and of both the polar and azimuth angles $(\theta,\varphi)$ (2DCH) \cite{Chen2016Phys.Rev.C44301} have been constructed. Here, for completeness, the frameworks of both the 1DCH and 2DCH are briefly given.

\subsubsection{One-dimensional collective Hamiltonian}

The 1DCH is written as \cite{Chen2013Phys.Rev.C24314, Chen2014Phys.Rev.C44306, Chen2016Phys.Rev.C54308}
\begin{equation}\label{eq-1DCH}
  \mathcal{H}(\varphi)= \frac{1}{2}B(\varphi)\dot{\varphi}^2 + \mathcal{V}(\varphi),
\end{equation}
in which $\mathcal{V}(\varphi)$ is the collective potential and $B(\varphi)$ is the mass parameter. The collective potential is obtained by minimizing the total Routhian $E'(\theta,\varphi)$ in the TAC with respect to $\theta$ for given $\varphi$, and the corresponding $B(\varphi)$ is calculated following Ref. \cite{Chen2013Phys.Rev.C24314}.

From the general Pauli prescription \cite{Pauli1933}, the quantal collective Hamiltonian reads
\begin{equation}\label{eq-1DCHCH}
  \hat{\mathcal{H}} = -\frac{\hbar^2}{2\sqrt{B(\varphi)}} \frac{\partial}{\partial \varphi} \frac{1}{\sqrt{B(\varphi)}}+\mathcal{V}(\varphi).
\end{equation}
The corresponding eigen energies $E^i$ and the wavefunctions $\Psi^i(\varphi)$ can be obtained by diagonalizing the Hamiltonian \eqref{eq-1DCHCH} via the basis expansion method, see Ref.~\cite{Chen2013Phys.Rev.C24314} for details. The collective Hamiltonian \eqref{eq-1DCHCH} is invariant under the transformation $\hat{P}_{\varphi}:~\varphi \rightarrow -\varphi$ \cite{Chen2013Phys.Rev.C24314}. The eigenvalues of $\hat{P}_{\varphi}$ are ``$+$" or ``$-$", depending on whether the state is symmetric or antisymmetric with respect to the transformation. Therefore, the eigenstates can be divided into two separate groups, i.e., $P_{\varphi}=+$ and $P_{\varphi}=-$ groups, and the eigen energies of the two groups can be labeled as $E^i_+$ and $E^i_-$, respectively.

\subsubsection{Two-dimensional collective Hamiltonian}

The 2DCH is written as
\begin{equation}\label{eq-2DCH}
 \mathcal{H}(\theta,\varphi)= \frac{1}{2}B_{\theta\theta}\dot{\theta}^2 +\frac{1}{2}B_{\theta\varphi}\dot{\theta}\dot{\varphi} +\frac{1}{2}B_{\varphi\theta}\dot{\varphi}\dot{\theta} +\frac{1}{2}B_{\varphi\varphi}\dot{\varphi}^2 + \mathcal{V}(\theta,\varphi),
\end{equation}
in which $\mathcal{V}(\theta,\varphi)$ is the collective potential, and $B_{\theta\theta}(\theta,\varphi)$, $B_{\theta\varphi}(\theta,\varphi)$, $B_{\varphi\theta}(\theta,\varphi)$, $B_{\varphi\varphi}(\theta,\varphi)$ are the mass parameters, and they can be obtained by the TAC calculations \cite{Chen2016Phys.Rev.C44301}.

From the general Pauli prescription \cite{Pauli1933}, the quantal collective Hamiltonian reads
\begin{align}\label{eq-2DCHCH}
  \hat{\mathcal{H}} =& -\frac{\hbar^2}{2\sqrt{w}}\left[  \frac{\partial}{\partial \varphi} \frac{B_{\theta\theta}}{\sqrt{w}} \frac{\partial}{\partial \varphi} - \frac{\partial}{\partial \varphi} \frac{B_{\varphi\theta}}{\sqrt{w}} \frac{\partial}{\partial \theta}  \right. \left. -\frac{\partial}{\partial \theta} \frac{B_{\theta\varphi}}{\sqrt{w}}\frac{\partial}{\partial \varphi} + \frac{\partial}{\partial \theta} \frac{B_{\varphi\varphi}}{\sqrt{w}} \frac{\partial}{\partial \theta} \right] + \mathcal{V}(\theta,\varphi),
\end{align}
in which $w$ is the determinant of the mass parameter tensor,
\begin{equation}\label{para_tensor}
  w=\det B = \begin{vmatrix} B_{\theta\theta} & B_{\theta\varphi} \\ B_{\varphi\theta} & B_{\varphi\varphi} \end{vmatrix}.
\end{equation}

The eigen energies $E^i$ and the corresponding wavefunctions $\Psi^i(\theta,\varphi)$ can be obtained by diagonalizing the Hamiltonian \eqref{eq-2DCHCH} via the basis expansion method, see Ref.~\cite{Chen2016Phys.Rev.C44301} for details. The collective Hamiltonian \eqref{eq-2DCHCH} is invariant under the transformation $\hat{P}_{\theta}:~\theta\rightarrow \pi-\theta$ or $\hat{P}_{\varphi}:~\varphi \rightarrow -\varphi$ \cite{Chen2016Phys.Rev.C44301}. The eigenvalues of $\hat{P}_{\theta}$ and $\hat{P}_{\varphi}$ are ``$+$" or ``$-$", depending on whether the state is symmetric or antisymmetric with respect to the transformations. Therefore, the eigenstates can be divided into four separate groups (${P}_{\theta}{P}_{\varphi}$), i.e., the positive-positive ($++$), positive-negative ($+-$), negative-positive ($-+$) and negative-negative ($--$) groups, and the eigen energies of the four groups can be labeled as $E^i_{++}$, $E^i_{+-}$, $E^i_{-+}$ and $E^i_{--}$, respectively.

\subsection{Electromagnetic transitions}

As the tilted angles $\theta$ and $\varphi$ are chosen as the collective coordinates in the collective Hamiltonian, the quantum fluctuations of the tilted angles are now considered in the frameworks of the 1DCH and 2DCH. Therefore, for EM transitions, it is natural to go beyond the semiclassical approximation in the TAC approach to include the
quantum fluctuation effects.

In the TAC, the EM transition probabilities are calculated as the expectation values of the corresponding operators $M1$ and $E2$ semiclassically \cite{Frauendorf1997Nucl.Phys.A131, Frauendorf2000Nucl.Phys.A115}
\begin{align}
  &B^{M1}_{\mathrm{TAC}}(\theta,\varphi)= \frac{3}{8\pi}\{ [-\mu_z\sin\theta_J+\cos\theta_J(\mu_x\cos\varphi_J+\mu_y\sin\varphi_J)]^2  +[\mu_y\cos\varphi_J-\mu_x\sin\varphi_J]^2 \},\label{M1_TAC}\\
  &B^{E2(I\rightarrow I-2)}_{\mathrm{TAC}}(\theta,\varphi)= \frac{15}{128\pi} \Big\{ \Big[ Q_{20}\sin^2\theta_J + \sqrt{\frac{2}{3}}Q_{22}(1+\cos^2\theta_J)\cos2\varphi_J \Big]^2 \notag \\
  &~~~~~~~~~~~~~~~~~~~~~~~~~ +\frac{8}{3}[Q_{22}\cos\theta_J\sin2\varphi_J]^2 \Big\}, \label{E2_TAC}\\
  &B^{E2(I\rightarrow I-1)}_{\mathrm{TAC}}(\theta,\varphi)=\frac{5}{16\pi}\Big\{\Big[ \sin\theta_J\cos\theta_J(Q_{22}\cos2\varphi_J-\sqrt{\frac{3}{2}}Q_{20})\Big]^2 +[\sin\theta_J\sin2\varphi_J Q_{22}]^2 \Big\},\label{E21_TAC}
\end{align}
in which the intrinsic magnetic moments $\mu_i=\sum_{\tau=p,n}(g_{\tau}-g_R)\langle j_{i(\tau)} \rangle$ with the $g$-factors $g_{\tau}$ ($g_R$) for valence nucleons (rotor) and the angular momentum components $j_{i(\tau)}$ of valence nucleons on the $i$ axis, and the intrinsic electric quadrupole tensors $Q_{20}=Q_{0}\cos\gamma$ and $Q_{22}=Q_{0}\sin\gamma/\sqrt{2}$ with the intrinsic electric quadrupole moment $Q_0$.

Note that the orientational angles $(\theta_J,\varphi_J)$ in Eqs. (\ref{M1_TAC})$-$(\ref{E21_TAC}) describe the orientations of the angular momentum $\bm{J}$ in the intrinsic frame, and are in general different from the tilted cranking angles $(\theta,\varphi)$ in the TAC. For given tilted cranking angles $(\theta,\varphi)$ in the TAC, the components of $\bm{J}$ are calculated by
\begin{equation}\label{J_k}
  J_k=\langle\hat{j}_k\rangle + \mathcal{J}_k\omega_k, ~~~~k=1,2,3,
\end{equation}
where the first term is from the valence particles and holes, and the second term from the rotor. The orientational angles $(\theta_J,\varphi_J)$ are defined as
\begin{equation}\label{theta_J}
  \tan \theta_J  = \frac{\sqrt{J_1^2+J_2^2}}{J_3},~\tan \varphi_J = \frac{J_2}{J_1}.
\end{equation}

In the TAC, the self-consistent solution is obtained by minimizing the total Routhian, in which the tilted cranking angles $(\theta,\varphi)$ are the same as the orientational angles $(\theta_J,\varphi_J)$. In such case, the EM transitions are calculated with $(\theta_J,\varphi_J)=(\theta,\varphi)$, and the contributions from other orientations are neglected. The effects of the quantum fluctuations on EM transitions will be considered in the frameworks of the 1DCH and 2DCH.

\subsubsection{EM transitions in the 1DCH}

In the 1DCH, the total Routhian $E'(\theta,\varphi)$ is minimized with respect to $\theta$ for a given $\varphi$, and the collective wavefunction $\Psi^i(\varphi)$ represents the amplitude of the collective state $i$ with azimuth angle $\varphi$. Hence, the EM transitions in Eqs. \eqref{M1_TAC}$-$\eqref{E21_TAC} only depend on azimuth angle $\varphi$. Therefore, the $M1$ and $E2$ transition probabilities in the 1DCH are
\begin{align}
  &B^{M1}_{\mathrm{1DCH}}  = \int_{-\pi/2}^{\pi/2}\mathrm{d}\varphi \sqrt{B(\varphi)} B^{M1}_{\mathrm{TAC}}(\varphi) |\Psi(\varphi)|^2, \label{M1_1DCH}\\
  &B^{E2(I\rightarrow I-2)}_{\mathrm{1DCH}} = \int_{-\pi/2}^{\pi/2}\mathrm{d}\varphi \sqrt{B(\varphi)} B^{E2(I\rightarrow I-2)}_{\mathrm{TAC}}(\varphi) |\Psi(\varphi)|^2,\label{E2_1DCH} \\
  &B^{E2(I\rightarrow I-1)}_{\mathrm{1DCH}} = \int_{-\pi/2}^{\pi/2}\mathrm{d}\varphi \sqrt{B(\varphi)} B^{E2(I\rightarrow I-1)}_{\mathrm{TAC}}(\varphi) |\Psi(\varphi)|^2.\label{E21_1DCH}
\end{align}

The angular momentum in the 1DCH is \cite{Chen2013Phys.Rev.C24314}
\begin{equation}\label{J_1DCH}
  J^{\mathrm{1DCH}}_{\mathrm{coll}} = \int_{-\pi/2}^{\pi/2}\mathrm{d}\varphi \sqrt{B(\varphi)} J_{\mathrm{TAC}}(\varphi) |\Psi(\varphi)|^2.
\end{equation}
Similarly, a quantal correction $I^{\mathrm{1DCH}}_{\mathrm{coll}} = J^{\mathrm{1DCH}}_{\mathrm{coll}}-1/2$ \cite{Frauendorf2000Nucl.Phys.A115} should be applied.

\subsubsection{EM transitions in the 2DCH}

In the 2DCH, the collective wavefunction $\Psi^i(\theta,\varphi)$ represents the amplitude of the collective state $i$ with polar and azimuth angles $(\theta,\varphi)$.

Similar to the 1DCH, the $M1$ and $E2$ transition probabilities in the 2DCH are
\begin{align}
  &B^{M1}_{\mathrm{2DCH}} =  \int_{0}^{\pi} \mathrm{d}\theta \int_{-\pi/2}^{\pi/2}\mathrm{d}\varphi \sqrt{w} B^{M1}_{\mathrm{TAC}}(\theta,\varphi) |\Psi(\theta,\varphi)|^2,\label{M1_2DCH}\\
  &B^{E2(I\rightarrow I-2)}_{\mathrm{2DCH}} =  \int_{0}^{\pi} \mathrm{d}\theta \int_{-\pi/2}^{\pi/2}\mathrm{d}\varphi \sqrt{w} B^{E2(I\rightarrow I-2)}_{\mathrm{TAC}}(\theta,\varphi) |\Psi(\theta,\varphi)|^2, \label{E2_2DCH}\\
  &B^{E2(I\rightarrow I-1)}_{\mathrm{2DCH}} = \int_{0}^{\pi} \mathrm{d}\theta \int_{-\pi/2}^{\pi/2}\mathrm{d}\varphi \sqrt{w} B^{E2(I\rightarrow I-1)}_{\mathrm{TAC}}(\theta,\varphi) |\Psi(\theta,\varphi)|^2, \label{E21_2DCH}
\end{align}
and the angular momentum in the 2DCH is \cite{Chen2016Phys.Rev.C44301}
\begin{equation}\label{J_2DCH}
  J^{\mathrm{2DCH}}_{\mathrm{coll}} = \int_{0}^{\pi} \mathrm{d}\theta \int_{-\pi/2}^{\pi/2}\mathrm{d}\varphi \sqrt{w} J_{\mathrm{TAC}}(\theta,\varphi) |\Psi(\theta,\varphi)|^2.
\end{equation}
A quantal correction $I^{\mathrm{2DCH}}_{\mathrm{coll}} = J^{\mathrm{2DCH}}_{\mathrm{coll}}-1/2$ \cite{Frauendorf2000Nucl.Phys.A115} is also applied.


\section{Numerical details}\label{numerical}

In the present calculations, a system with one $h_{11/2}$ proton particle and one $h_{11/2}$ neutron hole coupled to a triaxial rotor ($\gamma=-30^\circ$) is considered. The coupling coefficients in the single-$j$ shell Hamiltonian are taken as $C_{\pi}=0.25$ MeV for the proton particle and $C_{\nu}=-0.25$ MeV for the neutron hole. The moments of
inertia for irrotational flow are adopted with $\mathcal{J}_0 = 40~ \hbar^2/\mathrm{MeV}$. These numerical details are the same as those in Refs. \cite{Frauendorf1997Nucl.Phys.A131, Chen2013Phys.Rev.C24314, Chen2016Phys.Rev.C44301}. In the calculations of the EM transition probabilities, the effective $g$-factors are setting as $g_p-g_R=1$ and $g_n-g_R =-1$, respectively, and the electric quadrupole moment is taken as $Q_0 = 1.0$ eb. These assignments are the same in the calculations with the 1DCH, 2DCH, TAC, and PRM.


\section{Results and Discussion}\label{results}

In Ref. \cite{Chen2016Phys.Rev.C44301}, by taking the basis states under the periodic boundary condition and diagonalizing the collective Hamiltonian for given rotational frequencies, the collective energy levels and the wave functions obtained by the 2DCH have been compared with those obtained by the 1DCH. Meanwhile, the angular momenta and energy spectra calculated by the 2DCH have been compared with those by the TAC approach and the exact solutions of PRM. Here we follow the same 1DCH and 2DCH calculations in Ref. \cite{Chen2016Phys.Rev.C44301} and extend the discussion there to the intraband $M1$ and $E2$ transition probabilities.

\subsection{1DCH}\label{result_1DCH}

\begin{figure}[htbp]
  \centering
  \includegraphics[scale=0.6,angle=0]{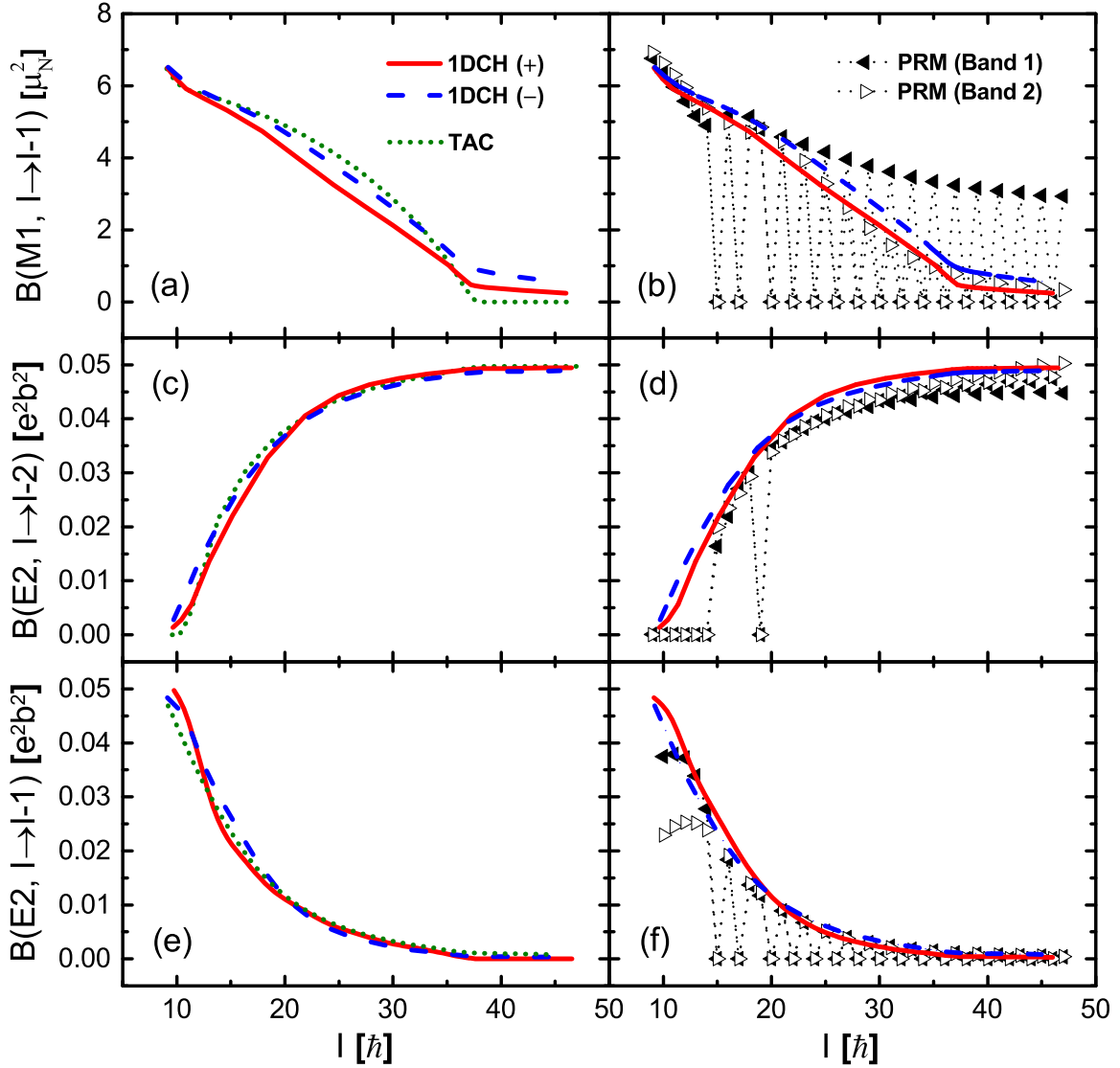}
  \caption{The intraband $M1$ and $E2$ transition probabilities of the doublet bands obtained by the 1DCH in comparison with the TAC [(a), (c) and (e)] and the PRM [(b), (d) and (f)] as functions of spin.}
\label{fig1}
\end{figure}

In Fig. \ref{fig1}, the intraband $M1$ and $E2$ transition probabilities of the doublet bands, i.e., the lowest states in the groups ($+$) and ($-$), obtained by the 1DCH in comparison with those by the TAC and the PRM as functions of spin are given.

In Figs. \ref{fig1} (a), (c) and (e), it is found that the tendencies of the $M1$ and $E2$ transition probabilities of the yrast band ($E^1_{+}$) in the 1DCH agree well with those in the TAC.
In Ref.~\cite{Frauendorf1997Nucl.Phys.A131}, it was shown that the TAC could reproduce the intraband transition probabilities for the yrast band in the PRM. The description of the chiral and wobbling excitations is beyond the mean field approximation in the TAC.
The 1DCH takes the quantum fluctuation of the azimuth angle $\varphi$ into account, and thus provides the intraband EM transition probabilities of both the yrast band ($E^1_{+}$) and side band ($E^1_{-}$). The obtained $M1$ and $E2$ transition probabilities for both bands are close to each other, as required by the chiral doublet bands or wobbling excitation bands.

\begin{figure}[htbp]
  \centering
  \includegraphics[scale=0.3,angle=0]{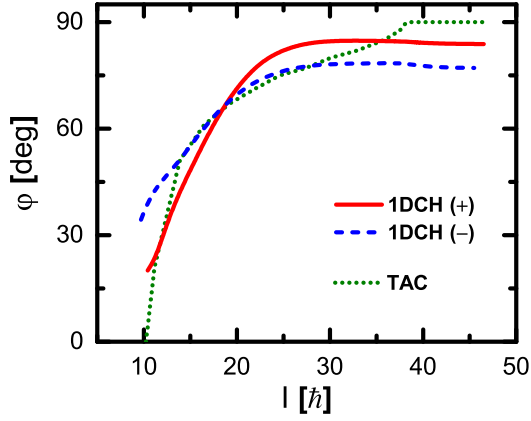}
  \caption{The effective azimuth angles $\varphi^{\mathrm{eff}}$ of in the doublet bands obtained by the 1DCH as functions of spin in comparison with the azimuth angle $\varphi$ by the TAC.}
\label{fig2}
\end{figure}

In Fig. \ref{fig1} (a), the $B(M1)$ values in the TAC drop rapidly to zero around $I= 37~\hbar$.
This is because the values of both polar and azimuth angles in the TAC become $\pi/2$ at this spin (see Figs. \ref{fig2} and \ref{fig4}), which means that the nucleus rotates with the intermediate axis. According to Eq. \eqref{M1_TAC}, the $M1$ transitions vanish.

In contrast, the $B(M1)$ values in the 1DCH approach to zero smoothly. This can be understood from the effective azimuth angles $\varphi^{\mathrm{eff}}$ in the 1DCH defined as
\begin{equation}\label{varphi_1DCH}
  \varphi^{\mathrm{eff}}_{\mathrm{1DCH}} = \int_{-\pi/2}^{\pi/2}\mathrm{d}\varphi \sqrt{B(\varphi)} |\varphi| |\Psi(\varphi)|^2.
\end{equation}
It is the expectation value of azimuth angle $|\varphi|$ including the quantum fluctuation effects of the orientational angles, and is displayed in Fig. \ref{fig2}.

Due to the quantum fluctuations, the orientation of angular momentum doesn't align with the intermediate axis at high spin but rather has a distribution. As a result, the effective azimuth angle $\varphi^{\mathrm{eff}}$ deviates from $\pi/2$, and the missing quantum effects in the TAC are resumed in the 1DCH. Therefore, the $B(M1)$ values in the 1DCH, although small, are non-vanishing at high spin.

In Figs. \ref{fig1} (b), (d) and (f), the intraband $M1$ and $E2$ transition probabilities in the 1DCH are compared with those in the PRM. For $B(E2,I\rightarrow I-2)$ values, both results of the yrast and side bands in the 1DCH agree well with those given by the PRM. For $B(M1)$ and $B(E2,I\rightarrow I-1)$ values, however, there is a noticeable difference between the PRM and the 1DCH. The results in the PRM present strong odd-even staggering behavior, whereas the ones in the 1DCH don't. The staggering behavior of the EM transitions of chiral doublet bands in the PRM has been analysed in Ref. \cite{Koike2004Phys.Rev.Lett.172502}. In the 1DCH, the angular momentum is not a good quantum number. Therefore, the staggering behavior, which strongly depends on the quantized angular momentum, is not reproduced in the 1DCH. Similar argument holds true for the TAC results where the staggering behavior can not be reproduced either. Nevertheless, it should be mentioned that the $B(M1)$ values in the PRM, regardless of the staggering behavior, are not exactly zero as well at high spin, in accordance with the results in the 1DCH.

\subsection{2DCH}\label{result_2DCH}

\begin{figure}[htbp]
  \centering
  \includegraphics[scale=0.6,angle=0]{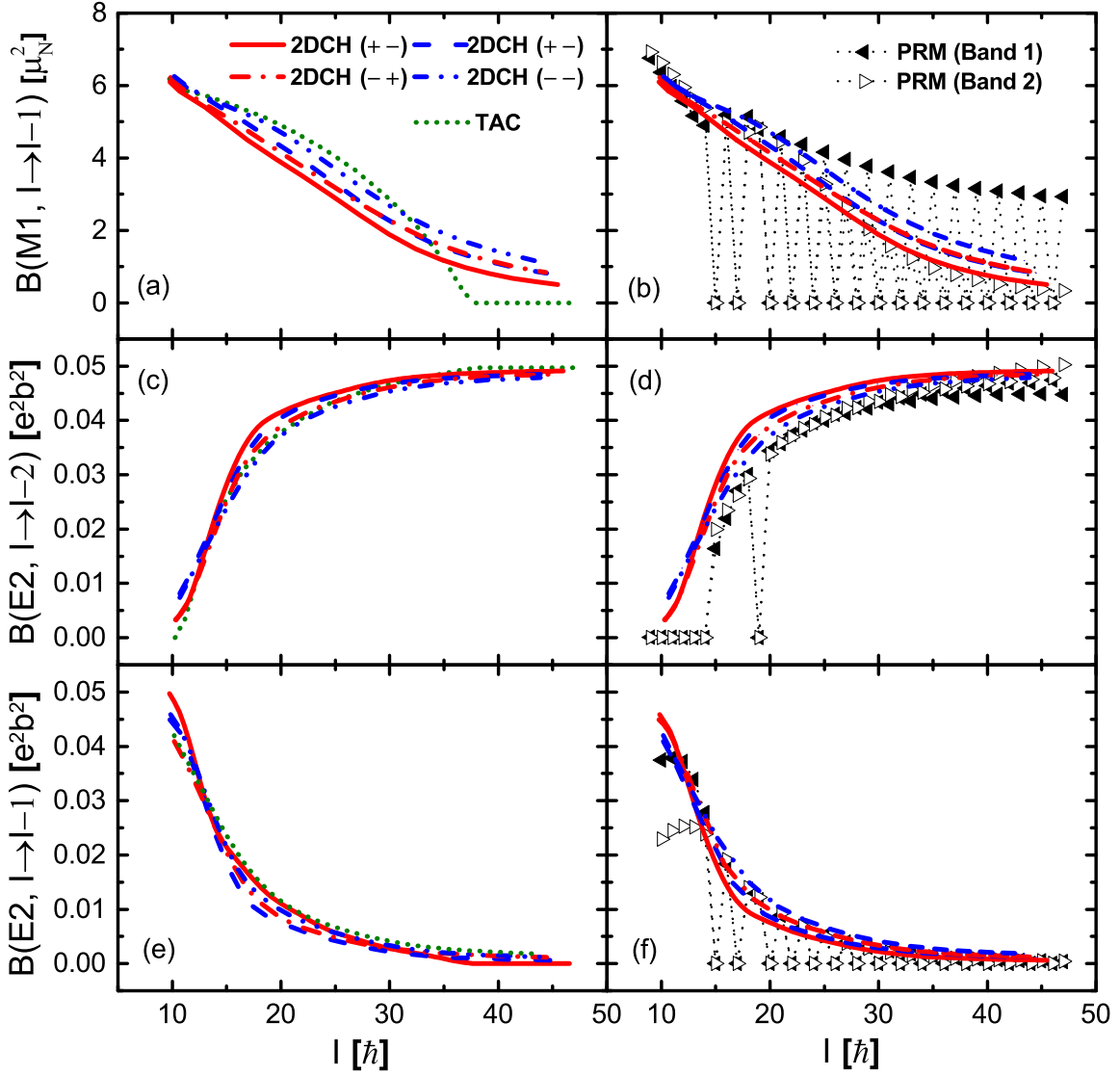}
  \caption{The intraband $M1$ and $E2$ transition probabilities of the lowest bands in the groups ($++$), ($+-$), ($-+$) and ($--$) obtained by the 2DCH in comparison with the TAC [(a), (c) and (e)] and the PRM [(b), (d) and (f)].}
\label{fig3}
\end{figure}

In Fig. \ref{fig3}, the intraband $M1$ and $E2$ transition probabilities of the lowest states in the groups ($++$), ($+-$), ($-+$) and ($--$) obtained by the 2DCH are compared with those by the TAC and the PRM.

In Figs. \ref{fig3} (a), (c) and (e), similar to the 1DCH, the tendencies of $M1$ and $E2$ transition probabilities of the yrast band ($E^1_{++}$) in the 2DCH agree well with those in the TAC. The $M1$ and $E2$ transition probabilities of the side bands ($E^1_{+-}$, $E^1_{-+}$, $E^1_{--}$) and the yrast band are close to each other, as required by the chiral doublet bands or wobbling excitation bands.

\begin{figure}[htbp]
  \centering
  \includegraphics[scale=0.3,angle=0]{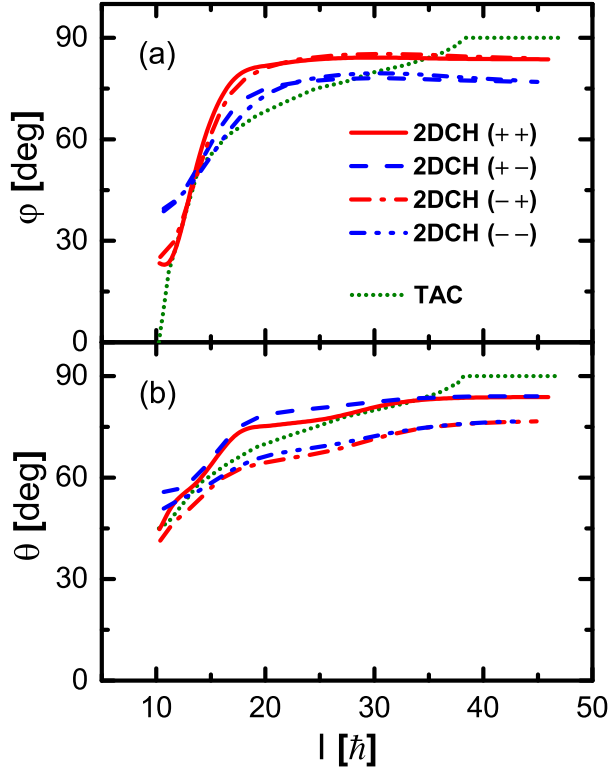}
  \caption{The effective tilted cranking angles $\varphi^{\mathrm{eff}}$ and $\theta^{\mathrm{eff}}$ of the lowest states in the groups ($++$), ($+-$), ($-+$) and ($--$) as functions of spin obtained by the 2DCH in comparison with the tilted cranking angles $\varphi$ and $\theta$ by the TAC.
  }
\label{fig4}
\end{figure}

In Fig. \ref{fig3} (a), the $B(M1)$ values in the 2DCH approach to zero smoothly at high spin, differing from the case in the TAC. Same as in the 1DCH, this can be understood from the effective azimuth angle $\varphi^{\mathrm{eff}}$ and polar angle $\theta^{\mathrm{eff}}$ respectively defined as
\begin{align}\label{effangle_2DCH}
  \varphi^{\mathrm{eff}}_{\mathrm{2DCH}} =& \int_{0}^{\pi} \mathrm{d}\theta \int_{-\pi/2}^{\pi/2} \mathrm{d}\varphi \sqrt{w} |\varphi|  |\Psi(\theta,\varphi)|^2,\\
  \theta^{\mathrm{eff}}_{\mathrm{2DCH}}  =& \int_{0}^{\pi} \mathrm{d}\theta \int_{-\pi/2}^{\pi/2} \mathrm{d}\varphi \sqrt{w} (\pi/2-|\pi/2-\theta|)  |\Psi(\theta,\varphi)|^2,
\end{align}
which are presented in Fig. \ref{fig4}.

At low spin ($I\leq 10~\hbar$), the azimuth angle $\varphi$ in the TAC is zero and the tilted cranking axis lies in the principal plane defined by the short- and long- axes, i.e., the so-called {\it planar solution} \cite{Frauendorf1997Nucl.Phys.A131}. However, in the 2DCH, the effective azimuth angles $\varphi^{\mathrm{eff}}$ are not zero, due to the quantum fluctuation effects of the orientational axis. Such quantum effects correspond to the {\it chiral vibrations} in the low spin region.

With increasing spin, the orientational axis does not lie in any of the principal planes in both the TAC and 2DCH. These are the so-called {\it aplanar solutions} \cite{Frauendorf1997Nucl.Phys.A131}, and they correspond to the {\it chiral rotation}. The values of $\theta^{\mathrm{eff}}$ and $\varphi^{\mathrm{eff}}$ in the 2DCH are close to but differ from $\theta$ and $\varphi$ in the TAC due to the quantum fluctuations in both the $\varphi$ and $\theta$ degrees of freedom.

At high spin ($I\geq 37~\hbar$), the tilted cranking axis in the TAC is along the intermediate axis. As a consequence, the $B(M1)$ value in the TAC drops to zero. However, in the 2DCH, the effective angles $(\theta^{\mathrm{eff}}, \varphi^{\mathrm{eff}})$ do not equal to $(\pi/2,\pi/2)$. Instead, the orientational axis has quantum fluctuations around the intermediate axis; corresponding to wobbling motions along $\theta$ and $\varphi$ directions, namely {\it $\theta$ wobbling} and {\it $\varphi$ wobbling} \cite{Chen2016Phys.Rev.C44301}. Therefore, as in the 1DCH, the $B(M1)$ values in the 2DCH are non-vanishing at high spin due to the quantum effects.

One remarkable feature in Fig. \ref{fig4} is that the effective angles $\varphi^{\mathrm{eff}}$ in the yrast band $E^1_{++}$ and the side band $E^1_{-+}$ are close to each other, whereas the $\theta^{\mathrm{eff}}$ in the yrast band $E^1_{++}$ and the side band $E^1_{+-}$ are close to each other. Similarly, the $\varphi^{\mathrm{eff}}$ in bands $E^1_{+-}$ and $E^1_{--}$ are close to each other, and the $\theta^{\mathrm{eff}}$ in bands $E^1_{-+}$ and $E^1_{--}$ are close to each other. This is because the states $E^1_{-+}$ and $E^1_{--}$ are one phonon vibrational states with $\theta$ respectively based on the states $E^1_{++}$ and $E^1_{+-}$. Similarly the states $E^1_{+-}$ and $E^1_{--}$ are the one phonon states with $\varphi$ respectively based on the states $E^1_{++}$ and $E^1_{-+}$. The $(\theta^{\mathrm{eff}},\varphi^{\mathrm{eff}})$ values for the yrast band ($E^1_{++}$) are almost the same as those for the side bands ($E^1_{+-}$, $E^1_{-+}$, $E^1_{--}$) around the spin $I= 15~\hbar$ in the 2DCH, which might be regarded as a signal for the static chirality in the 2DCH.

In Figs. \ref{fig3} (b), (d) and (f), the intraband $M1$ and $E2$ transition probabilities in the 2DCH are compared with those in the PRM. Again, the staggering behavior in the PRM can't be reproduced in the 2DCH, due to fact that the angular momentum in the 2DCH is not a good quantum number as discussed in Sec.~\ref{result_1DCH}. Except the staggering behavior, the amplitudes and tendencies of the $B(M1)$ and $B(E2)$ values in the PRM are reasonably reproduced by the 2DCH.


\section{Summary}\label{summary}
In summary, the intraband EM transition probabilities are calculated in the framework of collective Hamiltonian for chiral and wobbling modes. The EM transition probabilities for a system with one $h_{11/2}$ proton particle and one $h_{11/2}$ neutron hole coupled to a triaxial rotor ($\gamma=-30^\circ$) in the 1DCH and 2DCH are obtained and compared to the results given by the TAC and PRM.

The obtained EM transition probabilities for the yrast band and side bands in the 1DCH and 2DCH are close to those in the TAC. At high spin, the $B(M1)$ transition probabilities in the 1DCH and 2DCH have non-vanishing values as reflected by the effective orientational angles. This indicates that the missing quantum fluctuation effects of orientational axis are resumed.

The amplitudes and tendencies of the EM transition probabilities for the yrast and side bands obtained in the PRM can be well reproduced by the 1DCH and 2DCH. However, the odd-even staggering of the $B(M1)$ values can not be reproduced because the angular momentum is not a good quantum number in the 1DCH and 2DCH.

The successful descriptions of intraband EM transition probabilities here as well as the energy spectra in previous work \cite{Chen2016Phys.Rev.C44301} pave a road full of resplendent and magnificent prospect for building a collective Hamiltonian based on the microscopic tilted axis cranking covariant density functional theory \cite{Zhao2017Phys.Lett.B1} for chiral and wobbling modes.

\begin{acknowledgments}

X. H. Wu thanks Fangqi Chen for helpful discussions. This work was partly supported by the National Key R\&D Program of China (Contract No. 2018YFA0404400), the Deutsche Forschungsgemeinschaft (DFG) and National Natural Science Foundation of China (NSFC) through funds provided to the Sino-German CRC 110 ``Symmetries and the Emergence of Structure in QCD'', and the NSFC under Grants No. 11335002, No. 11621131001, and No. 11875075.

\end{acknowledgments}



\end{document}